\documentclass[%
 reprint,
 superscriptaddress,
 aps,
 prl,
]{revtex4-1}

\usepackage{amsfonts,amssymb,amsmath}
\usepackage{microtype}
\usepackage{txfonts}
\usepackage[unicode,breaklinks]{hyperref}
\usepackage{verbatim}
\usepackage{graphicx}
\usepackage{subfigure}
\usepackage{nicefrac}
\usepackage{placeins}

\newcommand{\beq}{\begin{equation}}
\newcommand{\eeq}{\end{equation}}
\newcommand{\beqa}{\begin{eqnarray}}
\newcommand{\eeqa}{\end{eqnarray}}

\usepackage{xcolor}
\usepackage{soul}
\usepackage[normalem]{ulem}

\begin{document}

\title{Propagation of first and second sound in a two-dimensional Fermi superfluid}

\author{A. Tononi}
\thanks{These authors have contributed equally.}
\affiliation{Dipartimento di Fisica e Astronomia ``Galileo Galilei'', Universit\`a di Padova, via Marzolo 8, 35131 Padova, Italy}

\author{A. Cappellaro}
\thanks{These authors have contributed equally.}
\affiliation{Dipartimento di Fisica e Astronomia ``Galileo Galilei'', Universit\`a di Padova, via Marzolo 8, 35131 Padova, Italy}
\affiliation{IST Austria (Institute of Science and Technology Austria), 
	Am Campus 1, 3400 Klosterneuburg, Austria}

\author{G. Bighin}
\thanks{These authors have contributed equally.}
\affiliation{Institut f\"ur Theoretische Physik, Universit\"at Heidelberg, Philosophenweg 19, D-69120 Heidelberg, Germany}
\affiliation{IST Austria (Institute of Science and Technology Austria), 
	Am Campus 1, 3400 Klosterneuburg, Austria}

\author{L. Salasnich}
\affiliation{Dipartimento di Fisica e Astronomia ``Galileo Galilei'', Universit\`a di Padova, via Marzolo 8, 35131 Padova, Italy}
\affiliation{Istituto Nazionale di Ottica (INO) del Consiglio Nazionale delle Ricerche (CNR), \\ via Nello Carrara 1, 50125 Sesto Fiorentino, Italy}
\affiliation{Istituto Nazionale di Fisica Nucleare (INFN), Sezione di Padova, 
via Marzolo 8, 35131 Padova, Italy}

\date{\today}


\begin{abstract}
Sound propagation is a macroscopic manifestation of the interplay between the equilibrium thermodynamics and the dynamical transport properties of fluids. Here, for a two-dimensional system of ultracold fermions, we calculate the first and second sound velocities across the whole BCS-BEC crossover and we analyze the system response to an external perturbation. In the low-temperature regime we reproduce the recent measurements [Phys Rev. Lett. {\bf 124}, 240403 (2020)] of the first sound velocity, which, due to the decoupling of density and entropy fluctuations, is the sole mode excited by a density probe. Conversely, a heat perturbation excites only the second sound, which, being sensitive to the superfluid depletion, vanishes in the deep BCS regime, and jumps discontinuously to zero at the Berezinskii-Kosterlitz-Thouless superfluid transition. A mixing between the modes occurs only in the finite-temperature BEC regime, where our theory converges to the purely bosonic results. 
\end{abstract}

\maketitle


{\it Introduction}.---
Investigations on the propagation of sound through a medium allow to test the microscopic theories on the structure of matter and to develop new theoretical ideas
\cite{Balibar:2007,Stringari:2016,Chester:1963,Schafer:2010,Schafer:2013,Zwierlein:2019,Volosniev:2020}.  
Along the historical development of physics, the concept itself of sound -- along with other physical entities -- has evolved and expanded to describe the new experimental evidence, refining our understanding of nature.

As a remarkable example of this process, we consider the propagation of sound in quantum liquids. The two-fluid theory of Tisza and Landau \cite{Tisza:1938,Landau:1941} explained the low-temperature experiments with $^4$He \cite{Kapitza:1938} describing it as a mixture of a normal (viscous) component and of a superfluid (non-viscous) one. 
The in-phase oscillation of these components, corresponding to the usual density wave and excited by a density perturbation, was denoted as the \textit{first} sound. The out-of-phase oscillation, corresponding to a heat wave and excited by a local heating of the fluid, was called the \textit{second} sound  \cite{Peshkov:1946,Lane:1947,Donnelly:2009}. 
This approximate description in which density and heat waves are decoupled holds for strongly-interacting superfluids like $^4$He and unitary Fermi gases  \cite{Ozawa:2014,Stringari:2015,Ota:2017}. However, it fails for weakly interacting quantum gases, where the isothermal and adiabatic compressibilities substantially differ \cite{Ozawa:2014}.  
In these systems, an experimental protocol consisting either of a density probe or of a heat one excites -- with different amplitudes -- both the first and the second sound: the sound modes are thus mixed (or hybridized) and the full solution of the Landau equation of sound is required. 

In uniform quantum gases, the richest phenomenology regarding sound propagation is offered by Fermi gases across the Bardeen-Cooper-Schrieffer (BCS) to Bose-Einstein condensate (BEC) crossover \cite{Heiselberg:2006}, in which the fermionic attractive interaction can be tuned from BCS weakly bound pairs to a BEC of composite bosons. 
Up to now, the experiments have mainly focused on 
three-dimensional fermions in cigar-shaped external potentials,
at unitarity \cite{Grimm:2013} and across the whole crossover \cite{Denschlag:2020}. 
As far as two-dimensional (2D) systems are concerned, a thorough theoretical description of sound propagation, including the physics of the  Berezinskii-Kosterlitz-Thouless (BKT) transition mediated by the unbinding of the vortex-antivortex dipoles 
\cite{Berezinskii:1972,Kosterlitz:1973,Nelson:1977}, is currently lacking. Understanding whether and how mixing of the sound modes occurs is particularly important to benchmark the recent \cite{Moritz:2020} and forthcoming investigations on 2D fermionic 
gases. 

Here we describe the propagation of sound modes across the two-dimensional BCS-BEC crossover, developing a
theoretical framework which relies on the 
beyond-mean-field equation of state and takes into account the pair fluctuations of the order parameter. 
Moreover, we consider the renormalization of the bare superfluid density due to the
screening of the interaction between quantized vortices. 
In the low-temperature collisional regime (for the noncollisional one see Refs. \cite{Dalibard:2018,Stringari:2018,Cappellaro:2018,Wu:2020}), the comparison with recent measurements \cite{Moritz:2020} of the first sound velocity shows a good agreement. 
Confirming the experimental outcome, we find that an excitation protocol consisting of a density probe excites almost exclusively the first sound, a clear signal of the decoupling of density and entropy modes across the whole BCS-BEC crossover.
This scheme changes slightly around the BKT critical temperature, where a partial mixing of the modes occurs in the BEC regime: we expect the hybridization to become more relevant as the system goes deeper into the BEC regime, reconnecting our theory to the framework of bosonic systems \cite{Ozawa:2014,Hadzibabic:2020}. 
We predict that a heat perturbation, due to the overall limited mixing, can easily excite the second sound: our results offer a solid benchmark for the future measurements of the velocity of second sound, which is an excellent and explicit probe of the BKT transition in uniform two-dimensional Fermi gases.


{\it First and second sound.}--- We consider a uniform two-dimensional superfluid at thermodynamic equilibrium.  
A local perturbation excites two wave-like modes -- the first and the second sound -- which propagate with velocities $u_1$ and $u_2$. Within the framework of Landau and Tisza two-fluid theory \cite{Tisza:1938,Landau:1941}, these velocities are determined by the positive solutions of the algebric biquadratic equation 
\begin{equation}
u^4 - ( c_{10}^2 + c_{20}^2 ) \ u^2 + c_T^2 c_{20}^2  =0 \; ,
\label{two-fluid-eq}
\end{equation}
namely, defining $u_1$ as the larger root and $u_2$ as the smaller one, 
\begin{equation}
\begin{aligned}
u_{1,2}^2 &= \frac{c_{10}^2 + c_{20}^2}{2} \pm \sqrt{\bigg(\frac{c_{10}^2 + c_{20}^2}
	{2}\bigg)^2 -c_{20}^2c_T^2} .
\end{aligned}
\label{solution in terms of cs and c2}
\end{equation}
Here we have introduced $c_{10}$, $c_T$, and $c_{20}$ as the adiabatic sound velocity, the isothermal and the entropic one, respectively: in specific thermodynamic regimes these velocities provide a good approximation and a clear physical interpretation of the sound modes $u_1$ and $u_2$. 
In particular, they read \cite{Heiselberg:2006}
\begin{equation}
c_{10}^2 = \bigg(\frac{\partial P}{\partial \rho}\bigg)_S \; ,
\qquad c_T^2 = \bigg(\frac{\partial P}{\partial \rho}\bigg)_T \;,
 \qquad c_{20}^2 = \frac{\rho_s T\,S^2}{\rho_n\,\rho \, L^2\,c_V}\;,
\label{sound velocity helium}
\end{equation}
where $P$ is the pressure, $S$ is the entropy, $\rho = \rho_s + \rho_n$ is the total
mass density, with $\rho_s$ ($\rho_n$) the superfluid (normal) mass density, respectively. Moreover, $c_V$ is the specific heat at constant two-dimensional volume $V=L^2$ (or area) of the system. 

In liquid helium and in unitary Fermi gases, where the approximate equality of the adiabatic and isothermal compressibilities implies that $c_{10} \approx c_T$ \cite{Ozawa:2014}, the sound modes of Eq.~\eqref{solution in terms of cs and c2} can be interpreted as a pure pressure-density wave and a pure entropy-temperature wave. 
The first sound, propagating with a velocity $u_1 \approx c_{10}$, is thus characterized by an in-phase oscillation of the superfluid and of the normal fluid, while, as a result of the out-of-phase oscillation of these components, the second sound propagates with a velocity $u_2 \approx c_{20}$. 

The simple picture of helium is no longer valid for Fermi gases in the deep BEC regime and for weakly interacting Bose gases, where the $c_{10} \approx c_T$ approximation breaks down due to the high compressibility of the system \cite{Ozawa:2014}. 
In this case, an external perturbation of the fluid induces a response in which the density-pressure and the temperature-entropy fluctuations are mixed. Then, according to the solution of Eq.~\eqref{solution in terms of cs and c2}, a
density probe, specified by a proper protocol, can excite both modes \cite{Denschlag:2020}. 
It is worth stressing that the current experiments with ultracold atoms can access both the amplitude and the velocity of propagating sound waves. In particular, if we consider the density response  
to an external perturbation, i.~e., $\delta\rho(r,t)$, the Landau
two-fluid model predicts 
$
\delta\rho(x,t) = W_1\,\delta\rho_1(r\pm u_1t) + W_2\,\delta\rho_2(r\pm u_2t)$, 
with $W_1$ the amplitude of the first sound mode and $W_2$ the amplitude of the second one \cite{Stringari:2015,Arahata:2009}. 
Here, the relevant experimental parameters are the relative amplitudes $W_{1,2}/(W_1+W_2)$, weighing the response of the system: these weights can be computed in terms of the sound velocities of Eqs.~\eqref{solution in terms of cs and c2} and \eqref{sound velocity helium} as \cite{Ozawa:2014,Nozieres-book}
\begin{equation}
\frac{W_{1}}{W_1+W_2} = \frac{(u_1^2 - c_{20}^2)\,u_2^2}{(u_1^2 -u_2^2)\,c_{20}^2},\; \; \frac{W_{2}}{W_1+W_2} = \frac{(c_{20}^2 - u_2^2)\,u_1^2}{(u_1^2 -u_2^2)\,c_{20}^2}\;.
\label{amplitude ratio}
\end{equation}
By definition, these complementary ratios add up to $1$, and the larger contribution among the two represents which mode is easier to detect by means of a density excitation protocol. 
In the following, after a microscopic derivation of the system thermodynamics, we will calculate the sound velocities $u_1$ and $u_2$ for two-dimensional uniform fermions across the whole BCS-BEC crossover.

\textit{Gaussian-pair fluctuations theory.}---
A mean-field description of a 2D fermionic gas is quantitatively accurate only 
in the BCS limit, and becomes extremely unreliable even in the intermediate interaction
regime. 
The order paramater fluctuations, neglected in the mean-field theory, are crucial to describe the full crossover at zero-temperature \cite{He:2015}, and particularly 
to recover the correct 
composite-boson limit in the deep-BEC regime \cite{Toigo:2015}. 
In this paper we adopt the Gaussian pair fluctuations (GPF) approach 
\cite{Engelbrecht:1997,Hu:2006,Diener:2008,Tempere:2012}, which 
has been also used to determine the bare \cite{Bighin:2016} and 
renormalized \cite{Bighin:2017} superfluid density in the 2D BCS-BEC crossover. 

A two-component 2D dilute Fermi gas can be described, in second quantization, by the Hamiltonian 
\begin{equation}
\begin{aligned}
\hat{H} &= \sum_{\sigma=\uparrow, \downarrow} \int_{L^2} 
\mathrm{d}^2 r \Big\{ \hat{\psi}^\dagger_\sigma (\mathbf{r}) 
\left( - \frac{\hbar^2}{2m} \nabla^2 - \mu \right) 
\hat{\psi}_\sigma(\mathbf{r}) +
\\
&+ g \ \hat{\psi}^{\dagger}_\uparrow (\textbf{r}) \
\hat{\psi}^{\dagger}_\downarrow (\textbf{r}) \
\hat{\psi}_\downarrow (\textbf{r}) \
\hat{\psi}_\uparrow (\textbf{r}) \Big\} \; ,
\end{aligned}
\label{bcs hamiltonian}
\end{equation}
where $\hat{\psi}_\sigma \left( \mathbf{r} \right)$ is the fermionic field operator 
which annihilates a fermion at position $\mathbf{r}$ with pseudo-spin 
$\sigma$. Here $m$ is the mass of a fermion and $g<0$ is the 
strength of the attractive contact interaction between atoms with opposite 
spins. 
The constraint $N = \sum_\sigma \int_{L^2} \mathrm{d}^2 r \ \langle 
\hat{\psi}^\dagger_\sigma (\mathbf{r}) \hat{\psi}_\sigma({\bf r}) \rangle$ imposes the conservation of the particle number $N$, 
and the interaction parameter $g$ can be related to the energy $\epsilon_\text{B}$ of a 
fermion-fermion bound state, see Ref.~\cite{Randeria:1989}.
To study the superfluid phase \cite{Stringari:2016}, one introduces the pairing field $\hat{\Delta}(\mathbf{r}) = g 
\hat{\psi}_\downarrow(\textbf{r}) \hat{\psi}_\uparrow(\textbf{r})$, corresponding
to a Cooper pair. 
In a mean-field treatment, the pairing field $\hat{\Delta}(\mathbf{r})$ is approximated with a constant real parameter, the pairing gap $\Delta_0$. 
This approximation leads to the mean-field thermodynamic grand potential
$\Omega_\text{mf} = \beta^{-1} \sum_{\mathbf{k}} \left[ \ln\{2 \cosh[\beta E_\text{sp}(k)]\} - \xi_k \right] - \Delta_0^2/ g$
with the usual definition of BCS fermionic elementary excitations  
$E_\text{sp}(k) = (\xi_k^2 + \Delta_0^2)^{1/2}$,
where $\xi_k  = {\hbar^2k^2 / 2m} - \mu$,
with $\mu$ the chemical potential and 
$\beta=1/(k_B T)$.

Building up on the mean-field theory just outlined, the two-dimensional nature of the system requires a better treatment, 
at least including the fluctuations of the pairing field up to the Gaussian 
level \cite{Engelbrecht:1997,Hu:2006,Diener:2008,Tempere:2012}. 
The Gaussian contribution 
to the grand potential, however, is considerably more involved, 
requiring several multidimensional integrations and the solution of non-trivial issues regarding 
regularization \cite{Salasnich:2016}. It reads:
$\Omega_\text{g} = (2 \beta)^{-1} \sum_{Q} 
\ln \det \mathbb{M} (Q) $
where $Q=(\mathbf{q},\mathrm{i} \Omega_j)$ and $\Omega_j = 2 \pi j / \beta$ 
are bosonic Matsubara frequencies, $j \in \mathbb{Z}$. 
The physics of the collective 
excitations is encoded in the matrix $\mathbb{M}$, the pair fluctuation 
propagator, whose matrix elements have involved analytical 
expressions, see Ref.~\cite{Engelbrecht:1997} for the explicit formulas. 
We derive the spectrum of bosonic collective excitations, i.e.~$E_\text{col}({\bf q}) = \hbar \omega({\bf q}) $, from the poles of the inverse pair fluctuation propagator, namely, by solving the equation $\mbox{det}(\mathbb{M}({\bf q},\omega)) = 0$. 
The total grand potential is then given by the sum of the mean-field and 
Gaussian contributions, $\Omega (\mu, T,L^2, \Delta_0) = 
\Omega_\text{mf} (\mu, T, L^2, \Delta_0) + \Omega_\text{g} 
(\mu, T, L^2, \Delta_0)$, 
from which it is possible to derive the gap equation, $\left( \partial \Omega_\text{mf}/\partial \Delta_0 
\right)_{\mu,T,L^2} = 0$, 
and the number equation,
$n = - L^{-2}
\left( \partial \Omega/\partial \mu\right)_{T,L^2}$,
with $n$ being the fermion density. Notice that the number equation 
is solved taking into account that $\Delta_0$ depends on 
$\mu$ \cite{Tempere:2012}. 

We derive the thermodynamic potential $\Omega$ by using, as input information,
the chemical potential $\mu$ and $\Delta_0$ from
the zero-temperature equation of state (EoS).  The temperature dependence of $\Omega$ is encoded
in the contributions related to the single-particle and pair fluctuation excitation spectra,
i.e., respectively, the first term in $\Omega_\text{mf}$ 
and the whole $\Omega_\text{g}$. 
We then evaluate the Helmholtz free energy as $F=\Omega(\mu,T,L^2,\Delta_0) + \mu N$, and, for an homogeneous system, the pressure reads $P = -\Omega(\mu,T,L^2,\Delta_0)/L^2$.
The entropy $S$ and the specific heat $c_V$ are calculated by differentiating $F$ with respect to the temperature, namely, $S =- (\partial_T F)_{L^2,N}$ and $c_V = -T(\partial^2_T F)_{L^2,N}$.
To calculate the adiabatic and isothermal velocities of Eq.~\eqref{sound velocity helium} we employ the following thermodynamical identity:  
$( \partial_{\rho} P  )_S
 = ( \partial_{\rho} P )_T + mN T/(\rho^2 c_V) [ ( \partial_{T} P )_{\rho}  ]^2$ \cite{Verney:2015},
where the derivatives of the pressure at the right-hand side can be evaluated applying the chain rule on $P=P(\mu,T,L^2,\Delta_0)$ and knowing $\mu$ and $\Delta_0$ from the EoS. 

\textit{Comparison with recent experiments.}---
The sound velocities of Eq.~\eqref{solution in terms of cs and c2} are a function of both the thermodynamical equilibrium properties discussed above and the superfluid density $\rho_s$ which,
instead, is a transport quantity. 
In two-dimensional systems, sound propagation is thus sensitive to the vanishing of $\rho_s$ at the BKT critical temperature $T_{\text{BKT}}$ \cite{Berezinskii:1972,Kosterlitz:1973,Babaev:1999}, where the thermally induced unbinding of the vortex-antivortex dipoles drives the system from the superfluid phase to the normal state. 

In the low-temperature regime of $T \ll T_{\text{BKT}}$, the superfluid density $\rho_s$ is very well approximated by the bare superfluid density $\rho_s^{(0)} = \rho - \rho_{n,F} - \rho_{n,B}$
(see Refs.~\cite{Bighin:2017,Fukushima:2007}) which includes two contributions to the normal density: $\rho_{n,F}$, of fermionic single-particle excitations whose spectrum $E_{\text{sp}}(k)$ is given above, and $\rho_{n,B}$ of bosonic collective excitations of the order parameter, described by $E_{\text{col}}(\mathbf{q})$.
Thus, following the Landau picture \cite{Landau:1941}, $\rho_s^{(0)}$ describes the superfluid depletion as driven exclusively by thermal excitations that, neglecting the contribution of the vortices, lead the system into the normal state at $T_c > T_{\text{BKT}}$.
In the temperature regime of $T \sim T_{\text{BKT}}$, due to the screening of the vortex-antivortex  interaction \cite{Kosterlitz:1973}, the 
bare superfluid density $\rho_s^{(0)}$ is renormalized to $\rho_{s}^{(R)}$. 
We calculate the renormalized superfluid density $\rho_{s}^{(R)}$ by jointly solving the Nelson-Kosterlitz renormalization group 
equations \cite{Nelson:1977} $\mathrm{d} K(\ell) / \mathrm{d} \ell = - 4\pi^3 K(\ell)^2 y(\ell)^2$ and
$\mathrm{d} y(\ell) / \mathrm{d} \ell =\left( 2 - \pi K(\ell) \right) y(\ell)$
for the running variables $K(\ell)$ and $y(\ell)$, where $\ell$ is the
adimensional scale. In the solution, we fix the initial conditions
$K(\ell = 0) = \beta J = \beta \hbar^2 \rho_s^{(0)}/(4m^2 )$ and 
$y(\ell = 0) = \exp(-\beta \mu_v)$, with $J$ being the phase stiffness
of the usual XY model, defined as $J= \hbar^2\rho_s^{(0)}/(4m^2)$ \cite{Jasnow:1973}
and $\mu_v = \pi^2 J/4$ being the vortex energy \cite{Zhang:2008}. 
Since the flowing stiffness displays a universal jump at the
transition, the renormalized superfluid density is given by $\rho_{s}^{(R)} = (4m^2 / \hbar^2) \ \beta \ K(\ell = +\infty)$.

\begin{figure}[hbtp!]
\centering
\hspace{-10pt} \includegraphics[width=1.025\columnwidth]{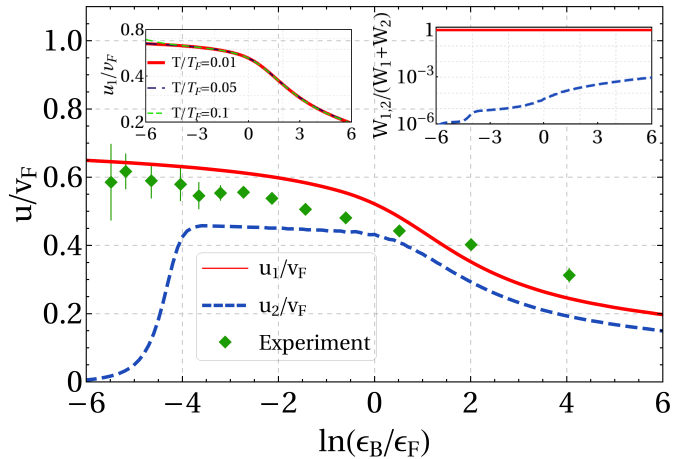}
\caption{
Evolution of the first sound velocity $u_1$ (red solid line) and of the second sound velocity $u_2$
(blue dashed line) along the BCS-BEC crossover,
calculated from Eq.~\eqref{solution in terms of cs and c2}.
The crossover is parametrized by $\ln(\epsilon_B/\epsilon_F)$, where
$\epsilon_F=\hbar^2\pi n/m$.
The sound modes are plotted at a fixed temperature $T/T_F = 0.01$, with
$T_F = \epsilon_F/k_B$, and the velocities rescaled in units of  
$v_F = \sqrt{2\epsilon_F/m}$. 
Measurements of the first sound \cite{Moritz:2020} (green points) at $T/T_F \le 0.1$ are in good agreement with our theoretical prediction, which is weakly dependent on temperature (see left inset).
Right inset: relative contribution to the density response of $u_1$ (red solid line) and $u_2$ (blue dashed line) [see Eq.~\eqref{amplitude ratio}] computed throughout the crossover at $T/T_F = 0.01$.
} 
\label{fig:1}
\end{figure}

\begin{figure}
\centering
\includegraphics[width=0.985\columnwidth]{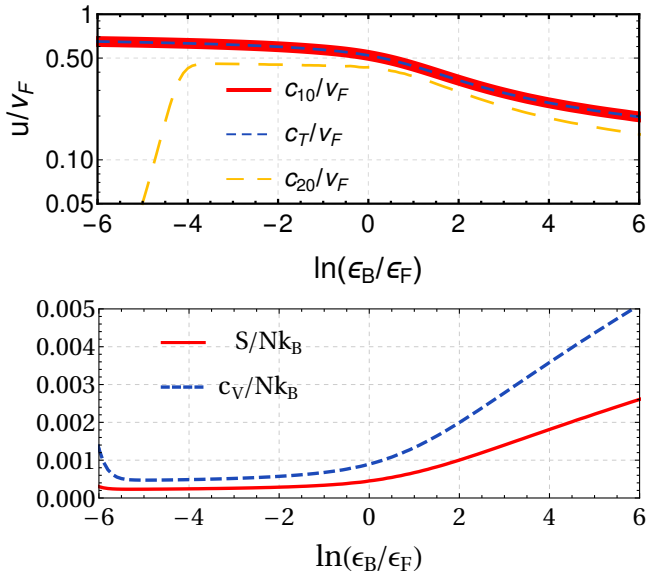}
\caption{
{ {	
Top panel: plots of the sound velocities defined in Eq.~\eqref{sound velocity helium}, as a function of the crossover parameter $\ln (\epsilon_B / \epsilon_F)$. 
Bottom panel: plots of the entropy per particle $S/(N k_B)$ and of the specific heat at constant volume per particle $c_V/(Nk_B)$, used to derive the low-temperature results of Fig.~\ref{fig:1}.
In both panels the temperature is fixed to $T/T_{\text{F}} = 0.01$.}}
}
\label{fig:3}
\end{figure}

\begin{figure*}[ht!]
\centering
\includegraphics[width=1.035\textwidth]{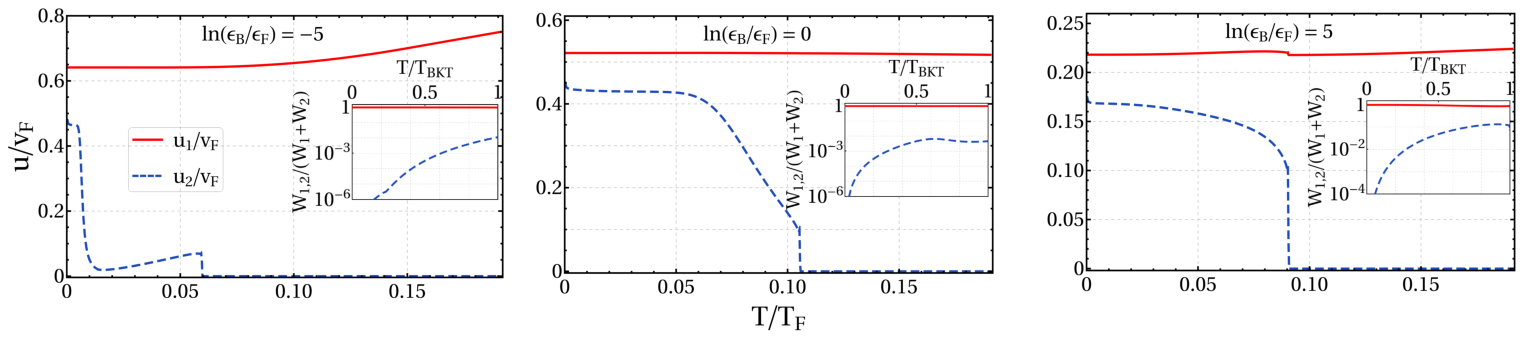}
\caption{	
First sound velocity $u_1 / v_F$ (red solid line) and second sound velocity $u_2 / v_F$ (blue dashed line) obtained from 
Eq.~\eqref{solution in terms of cs and c2}, plotted in terms of the rescaled
temperature $T/T_F$, for three different values of the crossover
parameter: $\ln(\epsilon_B/\epsilon_F)=-5$ (BCS regime), 
$\ln(\epsilon_B/\epsilon_F)=0$ (unitary regime),
and $\ln(\epsilon_B/\epsilon_F)=5$ (BEC regime).
The slower propagating mode $u_2$ disappears at the critical BKT temperature $T_{\text{BKT}}$. Insets: 
relative contribution to the density responses
 $W_{1,2}/(W_1+W_2)$ of $u_{1}$ (red solid line) and $u_2$ (blue dashed line) for the same 
 three values of the interaction parameter. 
}
\label{fig:2}
\end{figure*}

In Fig.~\ref{fig:1} we show the low-temperature behavior of the
sound modes of Eq.~\eqref{solution in terms of cs and c2}, where the 
thermodynamic functions have been derived from the Gaussian grand potential
$\Omega$ and the superfluid density is given by $\rho_s = \rho_s^{(R)}$.
The two sound velocities, $u_1$ (red solid line) and $u_2$ (blue dashed line), are 
displayed throughout the whole crossover, from $\ln(\epsilon_B/\epsilon_F) = -6$
(BCS side) to $+6$ (BEC side), at a fixed temperature of $T/T_F = 0.01$. 
The experimental points (green diamonds) are the measurements of the first sound velocity from Ref.~\cite{Moritz:2020}, and show a good agreement with our low-temperature theoretical prediction. 
The deviations from the theoretical curve could depend on the limited control of the temperature of the atomic ensemble, for which only an upper limit, $T/T_F \lesssim 0.1$, was provided \cite{Moritz:2020}\cite{private}. 
Moreover, our partial inclusion of thermal effects, i.e.~the use of the zero-temperature EoS for $\mu$ and $\Delta_0$ \cite{Bighin:2016}, could worsen the comparison with the experimental measurements in which the temperature was closer to the limit of $T= 0.1 \, T_F$.
Our theory also shows that, in agreement with the mean-field predictions of Ref.~\cite{Heiselberg:2006}, the velocity of the second sound $u_2$ vanishes for $\ln(\epsilon_B/\epsilon_F) \lesssim -5.5$, as a consequence of the superfluid depletion in the deep BCS limit. 

In the experiments, $u_1$ and $u_2$ are distinguished by measuring the amplitude of the two propagating modes \cite{Denschlag:2020}. 
In this regard, it is important to know in what proportion a density probe excites each mode, and 
in what regions of the crossover the observation of $u_1$ or $u_2$ is inhibited. 
In the two-fluid framework, this information is provided by the amplitude ratios of Eq.~\eqref{amplitude ratio}, shown in the right inset of Fig.~\ref{fig:1}. In the low-temperature regime discussed here and along the whole BCS-BEC crossover we find that $W_1/(W_1+W_2) \approx 1$ and that $W_2/(W_1+W_2) \approx 0$. These values of the ratios are a clear signal of the absence of mixing between the sound modes, and, therefore, that the sound velocities at low temperatures are well approximated by the expressions valid for liquid helium: $u_1 \approx c_{10}$ and $u_2 \approx c_{20}$. 
{ 
Note that the general solution of Landau equation of sound [see Eq.~\eqref{solution in terms of cs and c2}] gives these approximate equalities under the assumption that $c_{10} \approx c_T$. 
This condition is indeed verified in the low-temperature regime of $T/T_F = 0.01$, as can be seen from the top panel of Fig.~\ref{fig:3}, where we plot the adiabatic sound velocity $c_{10}$, the isothermal sound velocity $c_T$, and the entropic sound velocity $c_{20}$. 
}
Thus, as a prediction for the forthcoming experiments, we expect that a heat probe can easily and almost exclusively excite the second sound, for which we make a concrete quantitative prediction in Fig.~\ref{fig:1}. 
{ 
For the interpretation of future experiments, it is also useful to plot some relevant thermodynamic quantities calculated throughout our equation of state. 
In particular, in the bottom panel of Fig.~\ref{fig:3}, we plot the entropy per particle $S/(N k_B)$ and the specific heat at constant volume per particle $c_V/(N k_B)$. 

}

We have also calculated the spectral density function, whose Lorentzian peaks' width is 
related to the sound diffusion coefficient $D_s$. 
To effectively reproduce the results of the universal lower bound on $D_s$ \cite{Moritz:2020}, 
our theory should include higher order terms in the pairing field, especially on the BEC side of the 
crossover. Indeed, these terms are responsible of widening the spectral functions peaks, signaling that 
the collective excitations of the pairing field now have a finite lifetime
 \cite{Kamenev:2014,Bighin:2015}.

\textit{The role of {the BKT transition}.}---
The impact on the sound velocities of the BKT-driven renormalization of the superfluid density is clearly visible in Fig.~\ref{fig:2}, where, considering three different values 
of the crossover parameter $\ln(\epsilon_B/\epsilon_F)= \{-5, 0, +5 \}$, we plot $u_1$ and $u_2$ as a function of the temperature $T/T_F$. 
In every interaction regime, although with a different qualitative behavior, the mode $u_2$
disappears discontinuously at the critical temperature $T_{\text{BKT}}$. 
In addition, since due to the mixing both sounds depend on the superfluid density, $u_1$ also is discontinuous in the BEC regime, as can be seen in the right panel of Fig.~\ref{fig:2}. 
The jump of the first sound becomes more pronounced for larger values of $\ln(\epsilon_B/\epsilon_F)$, as one can expect from purely bosonic works \cite{Ozawa:2014,Hadzibabic:2020}, but here we limit to showing interaction regimes which can be conveniently reached in fermionic experiments ($\ln(\epsilon_B/\epsilon_F) \le 10$, see Ref.~\cite{Makhalov:2014}). 
We thus conclude that the discontinuities of the sound modes can probe the BKT transition in ultracold Fermi gases \cite{Dalibard:2018,Stringari:2018,Mathey:2020}. 
We also emphasize that, in the deep BEC limit, our theory provides a reasonable agreement with the BKT critical temperature obtained with purely bosonic theories, {as we discuss in the next section}.

In the insets of Fig.~\ref{fig:2} we report 
the relative contributions to the density response [see Eq.~\eqref{amplitude ratio}] whose general behavior is similar to that of the low-temperature regime, with a slight dependence on the interaction regime. Indeed, as before, the amplitude of the second sound $W_2/(W_1+W_2)$ is practically zero in the BCS regime and at unitarity. However, in the BEC regime the mixed response of the system emerges: $W_2/(W_1+W_2)$ increases with the temperature up to $0.15$, and jumps to zero in a sharp region around the critical $T_{\text{BKT}}$ temperature. 
Indeed, at $T>T_{\text{BKT}}$ only $u_1$ survives, corresponding to the standard propagation of sound in a normal fluid.

{ 
\textit{Composite boson limit.}---
In the deep-BEC limit the fermionic system can be mapped onto a system of interacting bosons with density $n_B = n_F/2$, 
mass $m_B = 2 m_F$ and chemical potential $\mu_B = 2 (\mu_F - \epsilon_B/2)$: 
the so-called `composite boson' limit; in this section we use explicit 'F' and 'B' subscripts to 
distinguish between bosonic and fermionic quantities. 
The bosonic and fermionic scattering lengths are related by the equation $a_B = \lambda \ a_F$ where $\lambda \approx 0.551$ \cite{Toigo:2015}. 
Therefore, the dimensionless coupling constant of 
a 2D Bose gas, $g_B$, is related to the fermionic quantities
 by the equation 
\begin{equation} 
g_B = - \frac{4\pi \hbar^2}{m_B} \frac{1}{2 \ln(k_F a_F) + \ln(\lambda^2 / 4 \pi)}\;
\label{mapping of g}  
\end{equation}
where 
\begin{equation}
\ln(k_F a_F) = \frac{1}{2}\bigg[-2\gamma + \ln\bigg(\frac{8\epsilon_F}{\epsilon_B}\bigg)\bigg]\;,
\end{equation}
with $\gamma \simeq 0.557$ being the Euler-Mascheroni constant. 
Another quantity which we need to map is the critical temperature $T_{\text{BKT}}^{(B)}$ of the system of composite bosons. 
As before, the superscript $B$ underlines that $T_{\text{BKT}}^{(B)}$ is the critical temperature of the Bose system
to which the fermionic system in the BEC side of the crossover can be mapped. 
To identify the temperature of the transition,
quantum Monte Carlo simulations \cite{Prokofev:2001,Bighin:2016} of 
2D Bose gases provide the universal relation
\begin{equation}
\frac{T^{(B)}_{\text{BKT}}}{T_F} = \frac{1}{2\ln\bigg[\dfrac{\xi}{4\pi}
	\ln\bigg(\dfrac{\pi}{e^{-2\gamma-1/2}}\,\dfrac{\epsilon_B}{\epsilon_F} \bigg)\bigg]}\;,
\label{critical temperature for composite bosons}
\end{equation}
with $\xi \simeq 554$ \cite{Bighin:2016}.

In current experimental setups the crossover parameter
can reach, at most, values around 
$\ln(\epsilon_B/\epsilon_F) \sim 10$ \cite{Makhalov:2014}. 
In this interaction range, the agreement between the bosonic theory and the composite boson limit  is not complete \cite{Bighin:2016}. 
We have verified it employing our finite-temperature theory for
$\ln(\epsilon_B/\epsilon_F) = 10$, which, according to Eq. \eqref{mapping of g}, corresponds to
the case of $g_B \simeq 1$ considered in Ref.~\cite{Ota:2017}. While the critical temperature is reasonably well
reproduced by our theory, the agreement of the sound velocities with the purely bosonic theory  is only qualitative.
}

\textit{Conclusions.}---   
We have calculated the first and the second sound velocities for a 2D Fermi gas across the BCS-BEC crossover, deriving the thermodynamics from the Gaussian pair fluctuations approach. 
Similarly to liquid helium, the second sound vanishes at the Berezinskii-Kosterlitz-Thouless temperature, where the superfluid component vanishes, heat propagation becomes purely diffusive, and the system supports only the usual (first) sound mode. 
At low temperatures, in accordance with recent experimental evidence, we do not observe the mixing of pressure-density oscillations and of entropy-temperature ones: a density probe excites only the first sound. Our theory reproduces the recently measured values of the first sound velocity and opens new experimental perspectives: we expect that a heat probe will excite only the second sound, for which we offer testable values and predictions, as a vanishing velocity in the deep BCS regime. 
We also discuss the thermal behavior of the sound modes, showing that, as can be expected from purely bosonic theories, a mixed response occurs only at finite temperatures and in the BEC regime, signaling the emergence of a bosonic character from composite bosons.

\begin{acknowledgments}
G.B.~acknowledges support from the Austrian Science Fund (FWF), under Project No.~M2641-N27.
This work was partially supported by the University of Padua, 
BIRD project ``Superfluid properties of Fermi gases in optical potentials."
The authors thank Miki Ota, Tomoki Ozawa, Sandro Stringari, Tilman Enss, Hauke Biss, Henning Moritz and Nicol\`o Defenu for fruitful discussions. The authors thank Henning Moritz and Markus Bohlen for providing their experimental data. 
\end{acknowledgments}


\begin{thebibliography}{99}
\bibitem{private} We fix $T/T_F=0.01$ to ensure that the BKT transition has not occurred for the whole range of $\ln(\epsilon_B/\epsilon_F)$ we consider, and that, therefore, the system is superfluid.
\end{thebibliography}
\end{document}